\documentclass[
 prb,
 amsmath,amssymb,
 aps
]{revtex4-2}
\usepackage{graphicx}
\usepackage{appendix}
\usepackage{bm}
\usepackage{braket}
\makeatletter
\newsavebox{\@brx}
\begin{document}

\title{Topological connections between the 2D Quantum Hall problem and the 1D quasicrystal}

\author{Anuradha Jagannathan}

\affiliation{Laboratoire de Physique des Solides, Universit\'{e} Paris-Saclay, 91405 Orsay, France}

\date{}

\begin{abstract}
1D quasicrystals such as the Fibonacci chain have been said to ``inherit" their topological properties from the 2D Quantum Hall problem. Yet, a direct way to see the connection was lacking until a common ancestor, the Fibonacci-Hall model, was introduced recently \cite{aj2025}. This 2D ancestor model relates the role of the external magnetic flux in the Hall problem and that of a geometric flux which describes the winding of the quasicrystal in 2D, in the cut-and-project method. Doing this enables us to extend the notion of Chern numbers as defined in 2D, to the energy bands of the 1D chain by adiabatic continuity. The older notion of gap labels in the 1D system are now seen to be derivable from the Chern numbers of the 2D bands. The Fibonacci-Hall model thus provides an important link between physics of two paradigmatic models, the Fibonacci quasicrystal and the quantum Hall insulator. The generalization to other 1D quasiperiodic models is expected to be relatively straightforward. The extension to 2D cut-and-project tilings is left for future studies. 
\end{abstract}

\maketitle
\section{Introduction}
This paper discusses the relation between two classes of condensed matter, both of which have led to prizes -- the Quantum Hall Effect (Physics Nobel prize 2015) and quasicrystals (Chemistry Nobel prize 2011). In both these areas, research has been intensively pursued since the 1980s. Although the two have overlapped for several decades \cite{bellissardpapers,bellissardpapers2}, an experimentally realisable  passage between the two problems has been proposed only rather recently, with the overall rise of interest in topological matter starting from the early 2000s. It is nowadays commonly accepted that the topological properties of the quasicrystal are ``inherited from" the 2D QH problem, as discussed in recent articles \cite{zilberberg1,zilberberg,zilberberg2}. It is the aim of this paper to describe one particular route by which the two problems can be connected. 

A pictorial representation of the two models is shown in Fig.1. The left figure illustrates a quasiperiodic 1D hopping model, involving two hopping amplitudes $t_a$ and $t_b$, for a segment of the Fibonacci chain. The right figure is a schema of the Quantum Hall (QH) problem of an electron hopping on square lattice and subjected to a perpendicular magnetic field. Each of the plaquettes is pierced by a magnetic flux, $\Phi$. The hoppings can be anisotropic, with an amplitude of $t_a$ for bonds oriented along the $x$-axis, and $t_b$ for bonds oriented along the $y$-axis, as shown in the figure. 

One way to establish the connection between the QH problem in 2D dimensions \cite{harper} and the Fibonacci quasicrystal is the following two step argument. The first step consists of rewriting the 2D Quantum Hall problem for an arbitrary dimensionless flux $\phi$ as an equivalent 1D problem, the Harper or Aubry-André model (AAH) \cite{harper,aamodel}. In the AAH model, the hopping amplitude along the 1D chain is uniform, while the onsite potential energies $V(x)$ vary as a cosine whose period is given by $\phi$. When the flux $\phi=\tau$ (where $\tau$ is the golden mean), the result is an incommensurate model whose spectral properties have been intensively studied, as has the connection between the QH problem and 1D quasicrystals \cite{bellissardpapers,bellissardpapers2}. The second step consists of noting, as  in \cite{kraus}, that the cosine potential can be continuously transformed into a discrete two-valued potential that follows the Fibonacci sequence. Arguing that this is an adiabatic transformation, they concluded that topological invariants of the model are conserved when going between the 1D models. More general forms of AAH models and their relation to quasicrystals have been investigated in a number of papers \cite{ganeshan,sanchez,strkalj}.

A different and more direct route to connect the 1D quasicrystal and the 2D QH problem has been recently proposed \cite{aj2025}. It uses the Cut-and-Project (CP) construction for the quasicrystal \cite{baake} to define a Fibonacci-Hall model which interpolates between the 2D and 1D problems. The principal feature in this construction is the introduction of the notion of ``geometric" flux. This quantity, which replaces the magnetic flux of the Quantum Hall effect, is an intrinsic characteristic of quasicrystals. It  derives from the winding property of the 1D selection strip in the 2D square lattice. In this paper, I explain the main ideas behind the Fibonacci-Hall model, and some of the main results obtained with it. By directly linking the 1D and 2D problems, the Fibonacci-Hall model provides useful physical intuition. It can also, particularly in higher dimensional cases, suggest new ideas for experimental measurements. In this context it is important to stress that topological invariants are not just a curious mathematical feature but determine measurable responses such as the quantum Hall conductivity -- which, due to its topological origin, is so incredibly precisely quantized that it is a standard for the resistance unit, the ohm. 

The paper is organized as follows: Sec.II presents basic definitions and results for bulk properties, and Sec.III presents some results on edge states in open chains. The final section presents a discussion of the model and some perspectives.

\begin{figure*}[h!]
\includegraphics[width=0.3\textwidth]{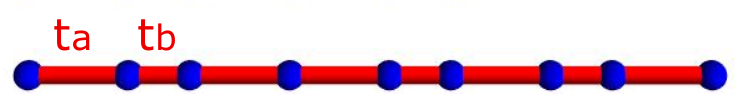} \hskip 2cm
\includegraphics[width=0.3\textwidth]{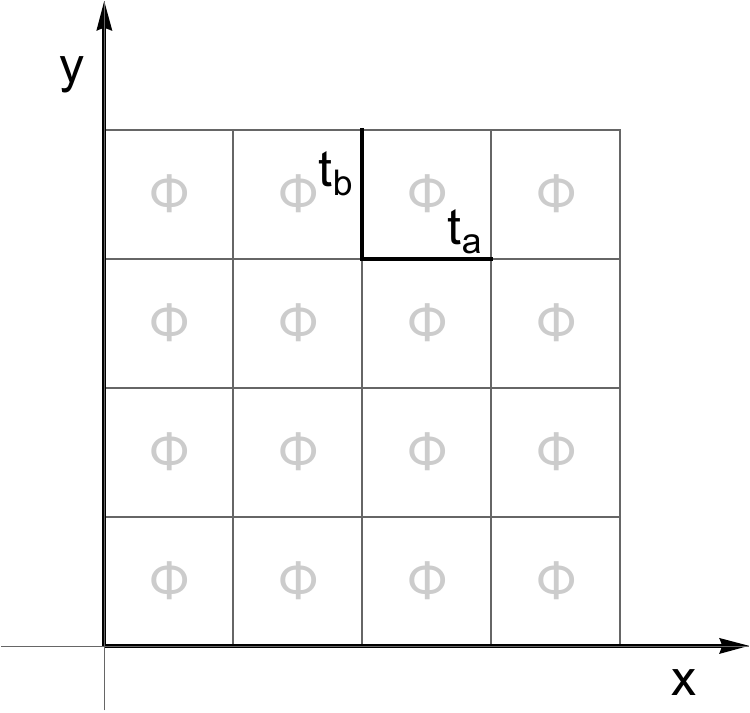} 
\caption{(Left) Schema for the Fibonacci hopping Hamiltonian with two different hopping amplitudes $t_a$ and $t_b$. (Right)  Schema for the Quantum Hall problem, showing a portion of a square lattice. The magnetic flux per plaquette of $\phi$ is shown, and hopping amplitudes $t_a$ and $t_b$ along the $x$ and $y$ directions respectively are indicated. }
\label{fig:qhmodel}
\end{figure*}

\section{The Fibonacci-Hall model}
The Fibonacci-Hall Hamiltonian $H_{FH}(\epsilon,\delta)$ describes a spinless electron hopping between sites of a square lattice, whose principal axes are defined to be the $x$ (horizontal) and $y$ (vertical) axes. Each of the square plaquettes of the lattice is pierced by a flux $\phi_S$ -- the geometric flux, which will be defined below. The vertical hopping amplitudes are proportional to $t_b$, and horizontal hopping amplitudes are proportional to $t_a$. 
The definitions below are given for finite approximants of the quasiperiodic chain. Thus, the Hamiltonian $H^{(k)}_{FH}(\epsilon,\delta)$ is $k$-dependent with the quasiperiodic limit reached when $k \rightarrow \infty$. For simplicity of notation, however, the index $k$ will be suppressed hereafter.

\begin{figure*}[h!]
\includegraphics[width=0.5\textwidth]{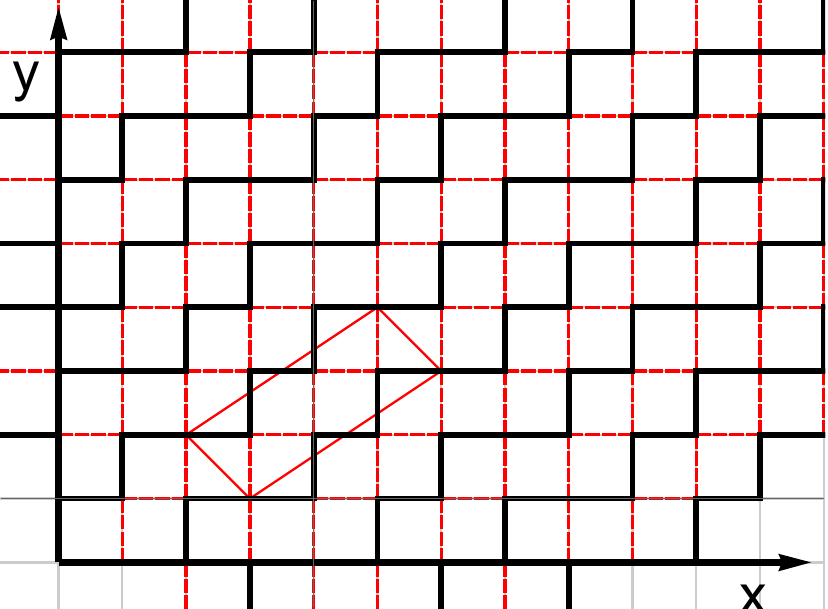} \hskip 2cm
\includegraphics[width=0.3\textwidth]{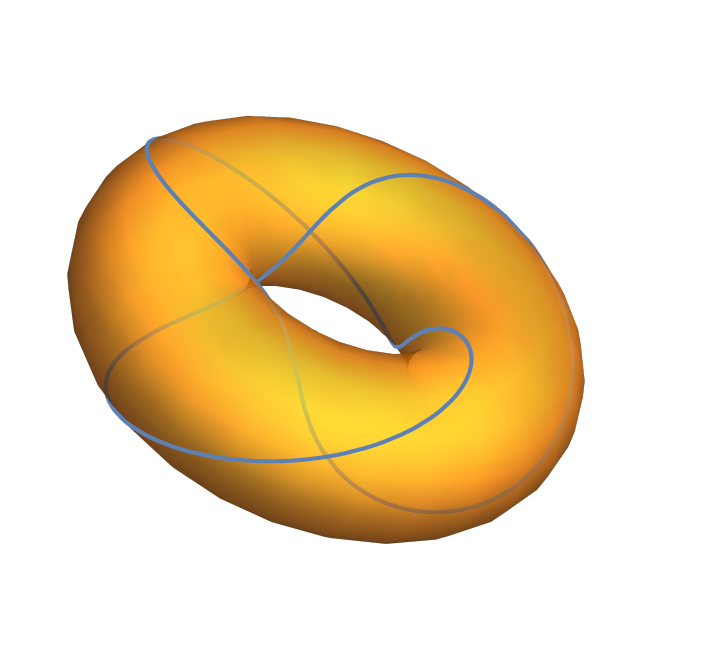} 
\caption{ (Left) A portion of the 2D plane illustrating the F-H Hamiltonian for approximant $k=4$ having 5 sites per unit cell. The  black zigzag lines join points belonging to different parallel approximant chains. Intra-chain hoppings take place on the thick black bonds while the inter-chain bonds are shown in red. A structural unit cell is outlined in red. (Right) A single unit square with edges identified to form a torus, showing the winding of the blue line of slope $2/3$ representing the strip S for the $k=4$ approximant.  }
\label{fig:unitcell}
\end{figure*}

We now recall the cut-and-project method used to generate approximants by selecting all the points lying within an infinite strip which is inclined with respect to the square lattice. The $k$th approximant is obtained for a strip with a rational slope equal to $F_{k-2}/F_{k-1}$, where $F_k$ is a Fibonacci number satisfying the recursion relation $F_k = F_{k-1}+F_{k-2}$ with $F_0=F_1=1$. The width of the strip corresponds to a unit square. It is easy to see that the entire 2D plane can be divided into parallel strips, as shown in Fig.2 (left). The points lying within a $given$ strip form an infinite zigzag line, namely, the $k$th approximant. It is periodic, with a number of sites per unit cell equal to $N=F_k$.

An important quantity entering the Fibonacci-Hall Hamiltonian is the geometric flux $\phi_S$. This is a dimensionless number related to the winding of the strip around the unit square. The number of turns for a general approximant is $F_{k-2}$ and $F_{k-1}$ along $y$ and $x$ directions respectively. The winding of the strip is shown in Fig.2 (right) for the approximant $k=4$ ($N=5$ sites). The choice $\phi_S= F_{k-1}/F_k$ incorporates this topological property, as well as a hierarchical structure present in the quasicrystal, as explained in more detail in \cite{aj2025}.

With these definitions in place, the Hamiltonian can now be written as
\begin{eqnarray}
    H_{FH} = \sum_{m,n} t^{y}_{mn} (c^\dag_{m,n}c_{m,n+1} + h.c.)   + t^{x}_{mn} (c^\dag_{m,n}c_{m+1,n} + h.c.) 
    \label{eq:fhham}
\end{eqnarray}
where sites are labeled according to their positions $\vec{r}_{mn}= ma\vec{x} + na\vec{x} $, where $a$ is the lattice constant. Just as in the original QH problem \cite{hofstadter}, it is convenient to work in the Landau gauge and take the fictitious vector potential parallel to the $y$ axis. With this choice, only the hopping amplitudes $t^y$ depend on $\phi_S$, while the hopping amplitudes $t^x$ are independent of the flux. Last but not least, we will distinguish between hopping within a given chain and hopping between different chains. The hopping amplitudes for inter-chain hopping are multiplied by a factor $\varepsilon$. The resulting set of hopping amplitudes are given in Table 1.

Fig.2 (left) shows a portion of the 2D plane illustrating the F-H Hamiltonian for the $k=4$ periodic approximant. Each of the slanted black zigzag lines represent one periodic approximant chain. Intra-chain hoppings take place on the thick black bonds while the inter-chain bonds are shown in red. The red parallelogram outlines a structural unit cell, while the magnetic unit cell is bigger, due to the extra phase factors introduced by the magnetic flux. Using the notion of magnetic translation operators (explained for example in \cite{tong}), and for a rational flux $p/q$ treated within the Landau gauge, the length of the magnetic unit cell is multiplied by $q$ along the $x$ direction only. For our tilted lattice, and for rational flux $=\frac{F_{k-1}}{F_k}$, a convenient choice for the magnetic unit cell is a parallelogram whose both sides are $F_k$ times longer than those of the structural unit cell.
  
\begin{table}[ht]
\caption{Definitions of the vertical and horizontal hopping amplitudes for interchain and intrachain hopping, with $t_b(n)= t_b \exp(2\pi i n \phi_S)$} 
\smallskip
\begin{center}
\begin{tabular}{llc}
     & inter-chain        & intra-chain           \\
     \hline
 $t^y_{mn}$ &$\varepsilon t_b(n)$ & $t_b(n)$          \\
$ t^x_{mn}$ &$\varepsilon t_a$ & $t_a$           \\
\hline
\end{tabular}
\end{center}
\end{table}

The F-H Hamiltonian depends on two parameters which can be varied, in addition to the fixed (for a given approximant $k$) flux parameter $\phi_S$. These parameters are: $\varepsilon$ such that $0\leq \varepsilon \leq 1$ and $\delta=t_b-t_a$ such that $-\infty< \delta < \infty$. The value of $t_a<t_b$, which determines only the overall energy scale is unimportant (and is set to $t_a=1$ in the numerical calculations). 
Periodic boundary conditions are imposed in the calculations of the electronic bands, taking care to use the magnetic unit cell, which has $N^3$ sites instead of $N$ for the structural unit cell.

Two of the limiting cases of the Fibonacci-Hall model for $\epsilon=0$ and  $\epsilon=1$ are well-known problems that have been extensively studied. For  $\varepsilon=0$ we have a set of identical Fibonacci approximants whose spectra and states are well understood for all values of $\delta$ (see for example the review in \cite{jagaRMP}). For $\epsilon=1$ we have the even more widely studied 2D QH Hamiltonian. When in addition $\delta=0$ one has the isotropic QH problem, corresponding to the celebrated Hofstadter butterfly \cite{hofstadter}. A third limit is the trivial case of decoupled periodic 1D chains, for $\varepsilon=\delta=0$. 

Outside of these limits, in the $\varepsilon-\delta$ plane, however, the behavior of this model is less well understood. A variety of topological phase transitions occur as parameters are varied. In particular, one observes several band crossings when $\delta\neq 0$ is fixed and $\varepsilon$ is varied. 
As a function of $\varepsilon$, one observes that levels of different bands of the F-H Hamiltonian cross, within each of the gaps. When $\delta=0$, in contrast, numerical studies indicate that gap crossings do not occur in any gap. To avoid gap crossings, therefore, we will define a path $\mathcal{P}$ to get from the 2D problem (Start) to the 1D Fibonacci approximant Hamiltonian corresponding to specified values of $t_b-t_a$ (Finish) as follows: \\
i) From the Start point $\varepsilon=1,\delta=0$,  reduce $\varepsilon$ down to a small finite value $\varepsilon_{min}$ \\
  ii) then increase $\delta$ to a value $\delta_{min}$ while decreasing $\varepsilon$ to 0 \\
iii) finally increase $\delta$ to its desired final value, to reach the Finish point .\\
Choosing this path avoids the origin, where all gaps close, and it allows for a continuous adiabatic transformation of the band structure, providing that $\varepsilon_{min}$ and $\delta_{min}$ are finite, and correspond to energy smaller than  the typical level spacing. Following this procedure, it was shown in \cite{aj2025} that the bands and gaps of the QH problem evolve smoothly to those of the Fibonacci approximant chain. The band Chern numbers therefore are invariant all along the path $\mathcal{P}$, until the 1D system is reached.

Chern numbers, in the QH problem, were shown to be fundamentally linked to an experimentally measurable quantity, the Hall conductance $\sigma_{xy}$ \cite{tknn}. On the one hand, a geometrical curvature in reciprocal space ($(k_x,k_y)$ expressed as dimensionless variables) of an insulating state corresponding to a filled band, $\psi$, can be defined as follows:
\begin{eqnarray}
    \mathcal{F}_{xy} =  \partial A_y/\partial k_x - \partial A_x/\partial k_y
\end{eqnarray}
where $A_\mu = -i \braket{ \psi \vert \partial/ \partial k_\mu \vert \psi } $ is the Berry connection. Thouless et al realized \cite{tknn} that the right-hand side in Eq.(1), when integrated over the Brillouin zone is essentially the Hall conductance (the so-called TKNN formula). In other words, if we define the Chern number $C$ by the integral of the curvature, 
\begin{eqnarray}
   C = -\frac{1}{2\pi}\iint_{BZ} \mathcal{F}_{xy}
\end{eqnarray}
then $\sigma_{xy}=(e^2/h) C$. Since, by its definition in Eq.2, the values of $C$ are quantized, the Hall conductance is quantized as well. When there are several bands, the conductance is proportional to the sum of the Chern numbers of bands below the gap, $\sum C_n$.  

In our model, the Chern numbers $C_n$ were calculated using a real space formalism introduced by Loring and Hastings \cite{loring} for Bott indices (which here are the same as their Chern numbers).  Knowing these values, one can deduce the following set of numbers for each of the $N-1$ gaps, 

\begin{eqnarray}
    g_n = \sum_{m\leq n} C_m
\end{eqnarray}
These $g_n$ are the gap labels discussed by Bellissard and coworkers \cite{bellissardpapers}. Even though the numbers $g_n$ do not correspond to a transport quantity, as in the 2D system, they can be accessed by measurements in open systems as discussed in the next section.

To illustrate our results, the computed Chern numbers for the bands and and gap of the $k=4$ approximant are given below in Table 2. Note the symmetry -- due to a chiral symmetry present in the hopping model, its energy bands are symmetric around $E=0$.

\begin{table}[ht]
\caption{Chern numbers and gap labels for the $N=5$ Fibonacci approximant  } 
\smallskip
\begin{center}
\begin{tabular}{cccccc}
\hline
$n$ & 1 & 2 & 3 & 4 & 5 \\
\hline
$\mathcal{C}_n$ & 2 & -3 & 2 & -3 &2  \\
$g_n$     &2 & -1 & 1 & -2 & -  \\ 
\end{tabular}
\end{center}
\end{table}

Thus, we see that the Fibonacci-Hall model allows one to deduce the band Chern numbers and the gap labels for the 1D chain by continuity. Now, we can compare them with a result on topological properties of aperiodic Schrodinger operators dating back to the 80s, namely the gap labeling theorem due to Bellissard \cite{bellissard}. This theorem concerns the value of the IDOS (integrated density of states), $I(E)$. Applied to Fibonacci models, the theorem states that, for an energy $E$ lying within a gap, $I(E)=p+q/\tau$ where $p$ and $q$ are two integers (the gap labels).  In the case of the Fibonacci approximants, this theorem can be modified \cite{maceicq} to describe the $k$th approximant chain. The IDOS in the $n$th gap  ($n=1,2,..,F_k-1$)  has a gap labeling as follows

\begin{eqnarray}
    I_n  = p+ q F_{k-1}/F_k \nonumber \\
  \rightarrow  n = p F_k + q F_{k-1}
    \label{eq:diophan}
\end{eqnarray}
where $p$ and $q$ are integers, with $\vert q\vert \leq N/2$. The second line, which follows from the relation $I_n = n/F_k$, is a Diophantine equation which can be solved to get $p$ and $q$. The integers $q$ are nothing but the gap labels $g_n$ defined above.

\begin{figure*}[h!]
\includegraphics[width=0.5\textwidth]{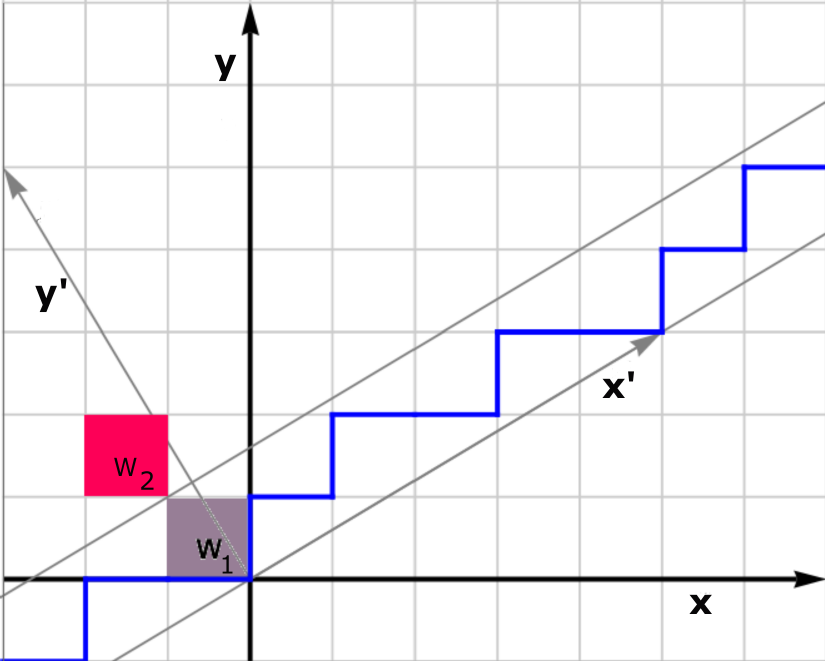} \hskip 0.5cm
\includegraphics[width=0.5\textwidth]{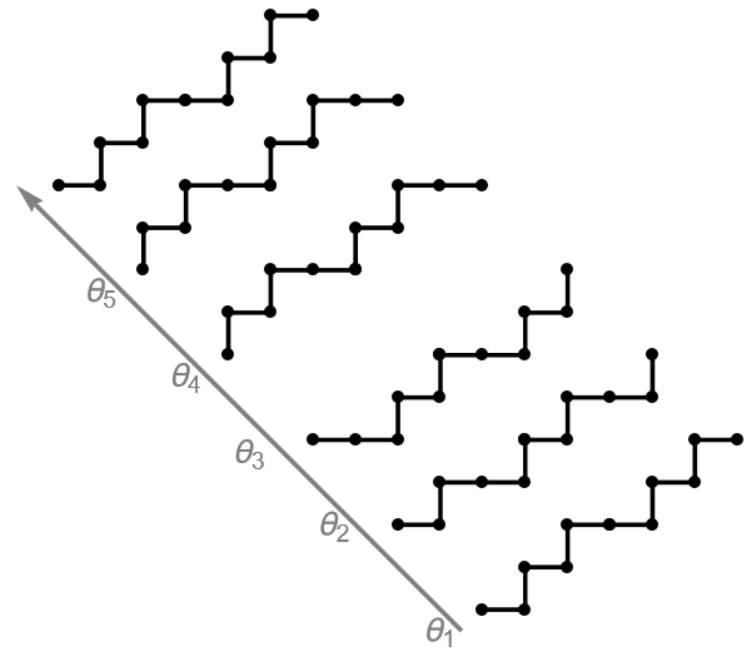}  \\
\caption{ (Left) The blue line joins points which lie within the CP selection strip shown with its window, $W_1$ (phase angle $\theta_1$). Shifting the window results in discrete changes of the blue zigzag line, which recovers its original shape (upto a shift) when the window reaches the position $W_2$ (phase angle $\theta_1+2\pi$) shown in red. The displacement corresponds to a complete cycle of the phase angle $\theta$. (Right) As the position of the selection window is varied and $\theta$ cycles through $2\pi$, the periodic approximant chain undergoes discontinuous changes. Successive changes are shown here for the case $N=5$. In each case only two unit cells of the infinite chains are shown. The original chain, shown at the bottom of the figure, changes discontinuously and cycles through 5 different shapes at $\theta=\theta_j$ ($j=1,...,5$), returning to its original shape at the end of the cycle. }
\label{fig:phasonflips}
\end{figure*}

As as we have already mentioned in the introduction, the fact that the gap labels are topological numbers has been known for some time, along with the connection between the Quantum Hall and the Fibonacci problems. Indeed, the gap labeling has been measured in careful laboratory experiments \cite{tanese,baboux}. The experiments rely on the properties of edge states, as we will explain in the next section.

To finish this section, some technical details are given. We have studied the Fibonacci-Hall model numerically for a set of small periodic approximants with number of sites ranging from $3$ to $13$ sites. Since the magnetic unit cells chosen here are a factor $N^2$ larger than the structural unit cell, the numerical diagonalization involves solving the eigenvalue problem for $N^3$ by $N^3$ complex sparse matrices. In these closed systems, these matrices additionally depend on the wave vectors $K$ and $K'$ arising due to periodic continuation in the 2D plane). The Chern number calculations do not show any effects of finite size, and results give exact values of the labels. On the other hand, the gap crossings in the phase space were seen to have a complex behavior as a function of parameters and system sizes, but this is left for a future study. All computations and plots were done using Mathematica.

\section{Open chains, edge states and phason angle $\phi$}
The central idea behind the laboratory measurements of topological invariants is based on the bulk-edge correspondence which states that there are edge states residing on the opened boundaries of a topological insulator. Although there are exceptions to this general statement, numerical studies show that this holds in the case of the Fibonacci chain \cite{prodan}. Opening the chain results in edge states, whose energies lie within the spectral gaps. Furthermore, there are expected to be precisely $\vert q\vert$ of them, where $q$ is the gap label.    

The experiments in \cite{tanese,baboux} were performed on Fibonacci approximant chains built from polaritonic resonators. The spatial intensity pattern of the resonant modes of the chain could be imaged and edge states could be identified. Their energies were measured and confirmed to lie within the spectral gaps. To count edge states, the experiments added an additional degree of freedom in the Hamiltonian, namely, the phason angle. To understand the role of the phason angle, we return to the CP representation of an approximant of the Fibonacci chain. The F-H Hamiltonian has a degree of freedom associated with the displacement of the window, a parameter $\theta$, the phason angle. Fig.3 (left) shows the strip of selected points corresponding to the selection window $W_1$. The phason angle describes shifts of the window along the diagonal of the unit square. Shifting from $W_1$ to the final position, $W_2$ corresponds to changing from $\theta=0$ to $\theta=2\pi$. The blue broken line evolves with $\theta$, undergoing successive phason flips, until it reaches its original shape after a complete cycle of $2\pi$. This is shown in Fig.\ref{fig:phasonflips} These phason flips do not modify the electronic properties of the periodic chains, however, they do so in finite chains. The differences induced by the shifts can be experimentally measured by following the changes of the far field diffraction pattern of waves incident on a finite sample of the quasicrystal \cite{dareau}. 

Varying $\theta$ results in modifications in the edge states. Their energies evolve with $\theta$, and move back and forth across the gap. One can count the number of gap crossings which occur for each of the gaps as $\theta$ varies between $[0,2\pi]$. Fig.3 (right) shows the evolution of the state energies as a function of $\theta$ for a $N=13$ approximant. The three biggest pairs of gaps have been labeled with their $q$ values. It can be seen that the edge state energies cross the respective band gaps $q$ times, and also that the direction of band crossings changes depending on the sign of $q$. This cycling in $\theta$ was the property using which the magnitude as well as the sign of the gap labels were experimentally measured.

\begin{figure*}[h!] 
\includegraphics[width=0.7\textwidth]{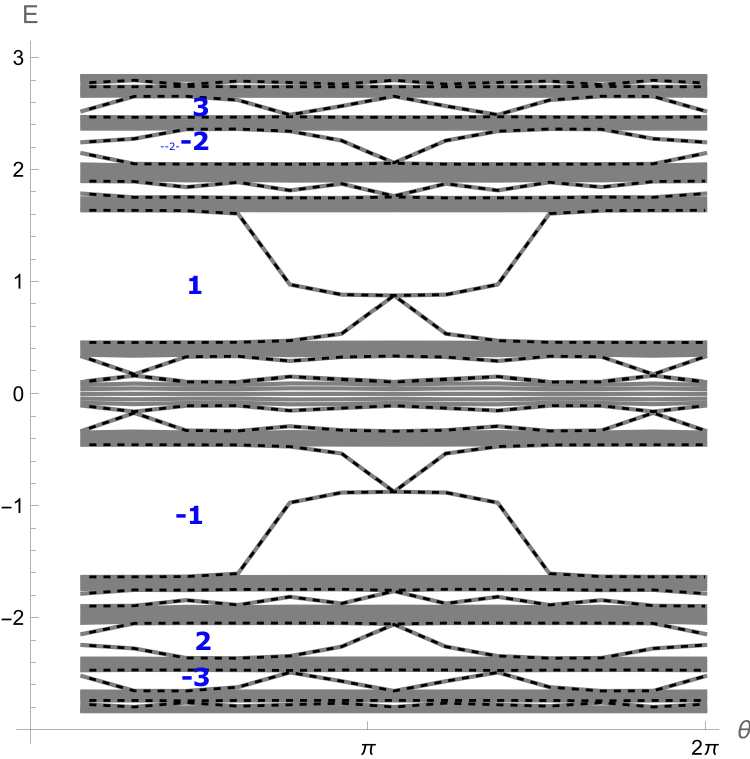}  
\caption{ Plot of energy bands and energies of edge states as a function $\phi$ for the $k=5$ approximant ($N=13$)}
\end{figure*}

In parallel with energy gap crossing, a charge transfer between the edges occurs as $\theta$ is varied. The edge states ``jump" across the chain $q$ number of times, as illustrated in Figs.4. The figures show the amplitude of the wave function on the sites of a periodic chain composed of blocks of $N=13$ approximant. The state considered lies at the top of the 2nd band, that is, just next to the second gap, whose gap label is $q=-3$. The figures show that, as the phase angle is varied through one complete cycle, the edge states lie on the left edge then the right edge and go back and forth $q$ times. This charge transfer occurs in the absence of any applied force, that is without an external electric field, and is thus termed ``topological". This phenomenon has been observed experimentally in optical waveguides \cite{kraus}, for a three band model. Topological charge transfer forms the basis for a recent proposal for a quantum quasicrystal ``pump" \cite{ghosh}.

Edge states of the quasicrystal can lead to many more interesting properties when additional couplings or interactions are added. When the quasicrystal is placed in contact with a superconductor, for example, there can be enhanced proximity induced superconducting order parameter due to edge states \cite{gautam}. Adding interactions in the Hamiltonian can give rise to new kinds of edge states, including Majorana fermions \cite{ghadimi,kobialka}.  

\begin{figure*}[h!]
\includegraphics[width=0.3\textwidth]{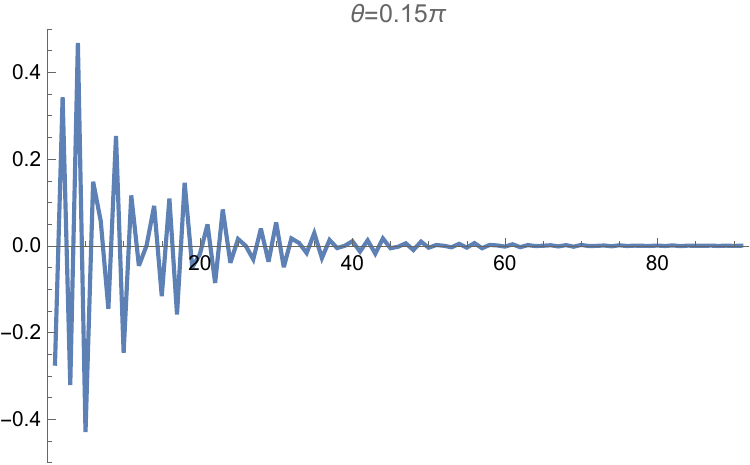} 
\includegraphics[width=0.3\textwidth]{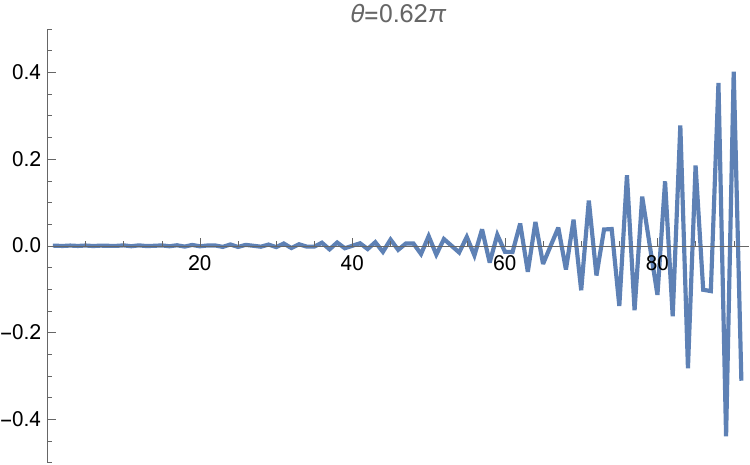} 
\includegraphics[width=0.3\textwidth]{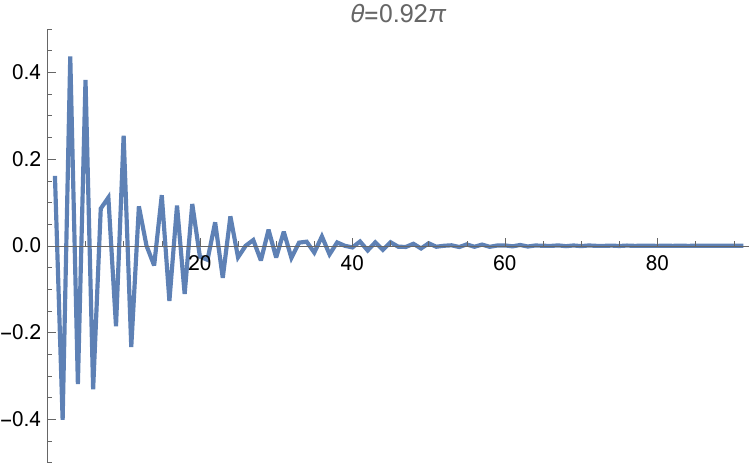}
\includegraphics[width=0.3\textwidth]{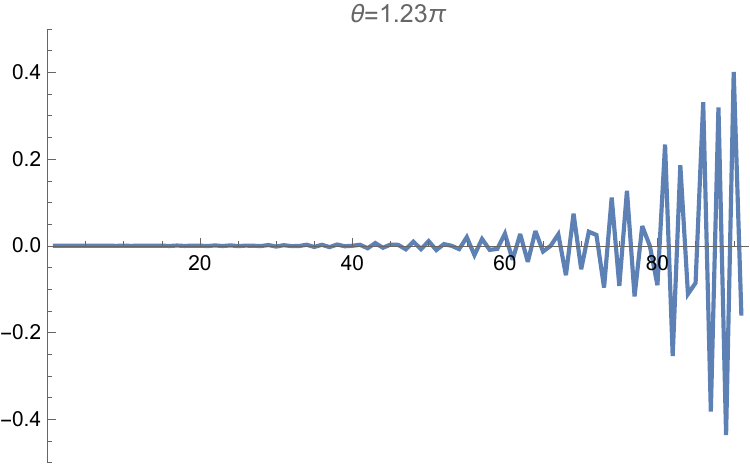}
\includegraphics[width=0.3\textwidth]{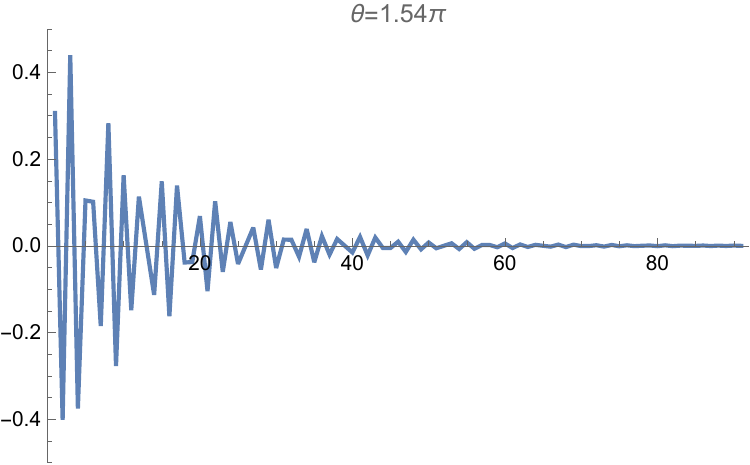}
\includegraphics[width=0.3\textwidth]{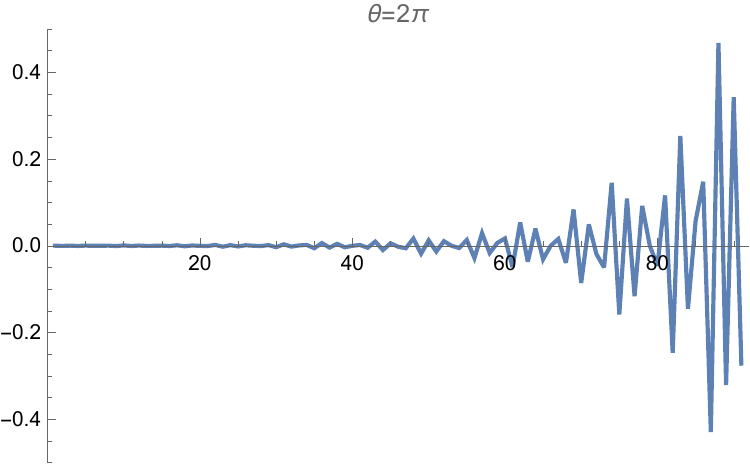}
\caption{ Plots of wavefunction amplitudes versus site number for a chain of 91 sites (7 unit cells of the $k=5$ approximant) showing the evolution of edge states as the phason angle $\phi$ is varied between 0 and $2\pi$. The edge states shown correspond to the energy gap of label $q=-3$. }
\end{figure*}



\section{Discussion}
We have shown, by introducing the Fibonacci-Hall model, that the 1D quasicrystal shares topological properties (Chern numbers) with the 2D Quantum Hall problem. The origin of non-trivial topology in the two models are the magnetic flux in the latter case, and the geometric flux due to the winding of the CP strip in the former. The bands of the 2D and 1D models can be related by a continuous transformation with no band crossings, along a particular path in the phase space of the model. In this way, while the Chern numbers are not defined for the 1D quasicrystal, one can get its 1D gap labels correctly by continuity from 2D. The topological meaning of the gap labels, which was already guessed and measured in experiments, is thereby explicitly confirmed. 
It is interesting to note that in an alternative scenario, Juricic et al \cite{juricic} have related topological invariants of the 1D system to those of the parent 2D Chern insulator, by computing local Chern numbers in an open system. The approach views quasicrystals as holographic images of higher-dimension topological crystals on lower-dimensional branes. 

We have discussed edge states and their evolution under changes of the phason angle. One of the striking phenomena that results, namely, topological charge transfer, was described.  In the 2D model, these edge states would be current-carrying and lead to the quantized Hall conductance. In the 1D system, they have been used to fabricate topological pumps in optical waveguides \cite{kraus}. Other applications, as in topologically protected qubits for quantum computing, have been envisaged.

Note that the 1D model discussed in this paper is the so-called off-diagonal model. This term refers to the fact that the quasiperiodic modulation in the Hamiltonian concerns only the hopping parameters, while all of the onsite energies are equal (and assumed to be equal to zero). The diagonal Fibonacci model, in which hoppings are uniform, but onsite energies are quasiperiodically modulated, belongs in the same topological sector. This model can be related to a 2D model in an way analogous to what we have presented here. It is also possible to extend the model so as to obtain topological invariants of other 1D quasicrystals such as the silver mean chain.

Finally, it will be interesting to explore whether the concepts used here could be generalized to higher dimensional quasicrystals such as the 2D Penrose or Ammann-Beenker tilings, which can be obtained by the CP method from 4D. Clearly, in higher dimensions, the topological structures are richer, leading to new physics and phenomena that could be theoretically and experimentally investigated. A first step in this direction has been taken with the realization of a synthetic 2D quasicrystal using cold atoms \cite{lohse}. In this pioneering experiment, a novel type of topological charge transfer, corresponding to quantized non-linear response under orthogonal electric and magnetic fields, was observed. It would of course be most interesting to devise an higher dimensional theory for models which are even closer to experimental alloy structures. Topological responses in realistic quasicrystals may be one of the most exciting fields of research in the future of this field.

\begin{acknowledgements}
I gratefully acknowledge many useful discussions with Frédéric Piéchon, Gilles Montambaux, Johannes Kellendonk and Pavel Kalugin.
\end{acknowledgements}



\bibliography{mybib} 

\end{document}